\documentclass[onecolumn]{revtex4}

\usepackage[dvipdfmx]{graphicx}
\usepackage{subfigure}
\usepackage{amssymb, amsmath,amssymb,amsfonts}
\usepackage{amsthm,mathrsfs,amsopn}
\usepackage{dcolumn}
\usepackage{bm}
\usepackage{color}
\usepackage[utf8]{inputenc}

\theoremstyle{plain} 

\newtheorem{theorem}{Theorem}

\newtheorem{remark}[theorem]{Remark}

\usepackage{cancel}

\begin{document}

\title{Turing patterns in a network-reduced FitzHugh-Nagumo model}

\author{ Timoteo Carletti$^1$}
\affiliation{$^1$naXys, Namur Institute for Complex Systems, University of Namur, Belgium}

\author{Hiroya Nakao$^2$}
\affiliation{$^2$Depatment of Systems and Control Engineering, Tokyo Institute of Technology, Tokyo 152-8552, Japan}

\begin{abstract}
{
Reduction of a two-component FitzHugh-Nagumo model to a single-component model with long-range connection is considered on general networks.
The reduced model describes a single chemical species reacting on the nodes and diffusing across the links of a multigraph with weighted long-range connections that naturally emerge from the adiabatic elimination, which defines a new class of networked {dynamical} systems with local and nonlocal Laplace matrices.
We study the conditions for the instability of homogeneous states in the original and reduced models and show that Turing patterns can emerge in both models.
}
\end{abstract}

\maketitle

\section{Introduction}
\label{sec:intro}

Self-organised structures arise in a large plethora of {natural} systems~\cite{prigogine,Murray}. Examples of emerging patterns are the spots and stripes on the coat or skin of animals~\cite{angelfish,leopard}, the architecture of large complex ecosystems~\cite{ecopattern} but also chemical and biological systems~\cite{Mikhailov,KapralShowalter,KeenerSneyd}. They are often the result of a combination of nonlinear reactions and local diffusion; in this framework the spatial self-organisation can be described following the visionary intuition of Alan Turing, who showed how nonlinear interactions between slow diffusing activators and fast diffusing inhibitors could induce patterns~\cite{turing,Meinhardt}. Pattern formation for systems evolving on regular lattices was further analysed by Othmer and Scriven~\cite{othmer} and then generalised to symmetric networks in Refs.~\cite{horsthemke,nakao}, {enabling to} better capture geometries present in many real cases.
Since then, a number of generalisations to various types of networks and dynamics has been made, including asymmetric, multiplex, time-varying, time-delayed, and non-normal networks~\cite{asllaniNatComm,asllaniPRE,kouvaris,Petit,Petit2,Muolo}, and intriguing relations between self-organisation and network topology have been revealed.

A key ingredient for the onset of Turing patterns is the much larger diffusivity of the inhibitor with respect to the one of the activator.
{In the first experimental realisation of Turing patterns by Castets {\it et al.}~\cite{castets}, a gel reactor was used to slow down the diffusion of the activator chemical species. The chemical mechanism for the slow activator diffusion that leads to formation of Turing patterns has been analysed by Lengyel and Epstein~\cite{lengyel} and more recently by Korvasov\'a {\it et al.}~\cite{korvasova}.
Mathematicians have also analysed the special case where the activator does not diffuse at all, described by ordinary differential equations coupled with a single reaction-diffusion equation, and showed that such a system can exhibit singular patterns~\cite{harting,anna}.}
{In models of biological systems, the effect of long-range inhibition has also been taken into account directly as nonlocal interaction rather than diffusive interaction~\cite{Murray,Kondo2,bresslof,Bullara,Cencetti}, which can also play the role of fast-diffusing inhibitor and lead to formation of Turing-like patterns.}

In this work, {we consider activator-inhibitor systems with slow-fast dynamics on general networks that undergo Turing instability.} We consider the limiting case of unbounded {reaction rate and} diffusivity for the inhibitor, {and approximately eliminate the dynamics of the fast inhibitor from the equations}; such approximation is known in the literature with the name of {\em adiabatic elimination}. For its relevance, we decided to perform our analysis on the networked FitzHugh-Nagumo model{, a typical mathematical model of activator-inhibitor systems with slow-fast dynamics~\cite{FitzHugh,Nagumo,Rinzel}}.
{Similar} adiabatic elimination has been already studied {by Ohta and collaborators} in the case of FitzHugh-Nagumo model defined {on} a continuous support~\cite{Ohta} {and has been used to analyse three-dimensional Turing patterns}~\cite{shoji}. {The aim of our study is to clarify the validity of such approximation on the pattern formation in networked systems and also to reveal its theoretical implications from the standpoint of dynamical processes on networks.}

{We show that,} under suitable assumptions, the adiabatic elimination process allow{s us} to reduce the system dimensionality and thus simplify the analysis of the model. In particular, we will prove the existence of Turing patterns for a networked reaction-diffusion model involving a single species, at odd with the known result stating the impossibility of such phenomenon {with purely diffusive interaction}~\cite{Murray}. Our result is grounded on the multigraph structure with long-range interactions of the network obtained using the adiabatic elimination; {indeed, even if the original system is defined on a simple network, the reduction process determines the emergence of a multigraph structure where several (weighted) links can connect two nodes}. The method we propose is general and thus its applicability goes beyond the model hereby considered.

The paper is organised as follows. The next section will be devoted to the introduction of the FitzHugh-Nagumo model and of the adiabatic elimination process. Then, in Section~\ref{sec:TP} we will describe the emergence for Turing patterns both in the original model and in the reduced one and we will compare the obtained results. We will then summarise and conclude our work. {Details of the calculations are given in the Appendices.}

\section{The model}
\label{sec:model}
Let us consider the FitzHugh-Nagumo model~\cite{FitzHugh,Nagumo,Rinzel} (for short {\em FHN} in the following):
\begin{equation}
\begin{cases}
\partial_t u(t,x) &=\mu u-u^3-v+D_u \nabla^2 u\\
\partial_t v(t,x) &=\gamma (u-\alpha v - \beta)+D_v \nabla^2 v\, ,
\end{cases}
\label{eq:FHNcont}
\end{equation}
where the model parameters, $\alpha$, $\beta$, $\gamma$, $D_u$ and $D_v$, are assumed to be positive, and the functions $u$ and $v$ to depend on time $t$ and on the spatial position $x$ on the domain of interest, where the latter can thus be a scalar or a vector according to the system dimensions. The variable $u$ promotes the growth of itself and of $v$ and can thus be named {\em activator}, on the other hand $v$ plays the role an {\em inhibitor}.

This model has been used to describe the impulse propagation along a nerve axion~\cite{Rinzel} {and also for analysing pattern formation in reaction-diffusion systems~\cite{Ohta,shoji}.} It is thus natural to consider its extension to a networked support {and consider the} possibility for the involved quantities $u$ and $v$ to diffuse across the network using the available links, and to interact only on the nodes. In this framework, the model~\eqref{eq:FHNcont} rewrites thus
\begin{equation}
\begin{cases}
\dot u_i(t) &=\mu u_i-u_i^3-v_i+D_u \sum_{j=1}^N L_{ij}u_j\\
\dot v_i(t) &=\gamma (u_i-\alpha v_i - \beta)+D_v  \sum_{j=1}^N L_{ij}v_j\, ,
\end{cases}
\label{eq:FHNnet}
\end{equation}
{for all $i=1, ..., N$}, where {we consider a connected and symmetric network consisting of $N$ nodes, $u_i$ and $v_i$ are the concentrations of the activator and inhibitor on node $i$, respectively}, $\mathbf{L}$ is the $N\times N$ Laplace matrix of the network defined by $L_{ij}=A_{ij}-k_i\delta_{ij}$, being $\mathbf{A}$ the adjacency matrix encoding the network connections, i.e. $A_{ij}=1$ if and only if there is a link connecting nodes $i$ and $j$. The number of connections of a given node is named the node degree, $k_i=\sum_j A_{ij}$. Let us observe that we are considering undirected networks whose matrices $\mathbf{L}$ and $\mathbf{A}$ are symmetric, however the extension to directed networks is straightforward (see {Appendix}~\ref{sec:dirnet}).
{The Laplace matrix $\mathbf{L}$ has a single $0$ eigenvalue $\Lambda^1$ associated with a uniform eigenvector $\boldsymbol{\phi}^1 \propto (1,\dots,1)^T$, and all remaining eigenvalues $\Lambda^{s}$, $s=2, \dots, N$ are negative.}

{We assume that the} species $u$ and $v$ spread with very different diffusion coefficients, say $D_v\gg D_u$. It is thus interesting to consider the limit {$\gamma \to \infty$ and} $D_v\rightarrow \infty$ {while keeping the ratio $\gamma / D_v$ constant,} {and adiabatically eliminate the variable $v_i$}. It has been proposed for the first time in~\cite{Ohta} for the {\em FHN} model on continuous support, Eq.~\eqref{eq:FHNcont}, while, to the best of our knowledge, it has never been applied on the networked version of the {\em FHN} model, Eq.~\eqref{eq:FHNnet}. The advantage{s} of this process are {(i) it} reduces the dimensionality of the model and thus somehow simplify the analysis {and (ii) it provides us with an alternative viewpoint on the self-organisation of Turing patterns on networks.} In the following, we will be interested in determining the onset of Turing patterns on the reduced {\em FHN} model and {compare the results with} the same phenomenon in the original model. 

Let us now explicitly compute the adiabatic elimination for the networked {\em FHN} model. System~\eqref{eq:FHNnet} can be cast in the slow-fast framework by defining $c=\gamma/D_v>0$ such that $c$ remains constant once $D_v$ grows unbounded, and thus $\gamma$ does. In this case, the second equation in~\eqref{eq:FHNnet} rewrites
\begin{equation*}
0 =c (u_i-\alpha v_i - \beta)+ \sum_j L_{ij}v_j\, 
\quad\forall i=1,\dots N\, ,
\end{equation*}
namely
\begin{equation*}
(\mathbf{L}-c\alpha \mathbf{I})\vec{v}=-c (\vec{u} - \beta)\, ,
\end{equation*}
being $\mathbf{I}$ the $N\times N$ identity matrix, {$\vec{u} = (u_1, ..., u_N)^T$, and $\vec{v} = (v_1, ..., v_N)^T$}.

Let us introduce the matrix $\mathbf{M}=(\mathbf{L}-c\alpha \mathbf{I})^{-1}$ {and represent $\vec{v}$ as $\vec{v} = - c \mathbf{M} (\vec{u} - \beta)$}.
We here observe that the spectrum of $\mathbf{L} -c\alpha \mathbf{I}$ is always formed by negative eigenvalues because $c\alpha >0$ and the matrix $\mathbf{M}$ is thus well defined.
{Plugging this $\vec{v}$ into the equation for $\vec{u}$,} the remaining equation of~\eqref{eq:FHNnet} becomes
\begin{equation}
\label{eq:fhnmodel}
\dot{u}_i = \mu u_i-u_i^3+c\sum_{j}M_{ij}(u_j-\beta)+D_u \sum_j L_{ij}u_j
\quad\forall i=1,\dots N\, .
\end{equation}
Let us observe that if {we denote by} $(\Lambda^s,\boldsymbol{\phi}^s)$, $s=1,\dots,N$, a set of orthonormal eigenvalue -- eigenvector couples for $\mathbf{L}$, that is, $\sum_i \phi^s_i\phi^{s'}_i=\delta_{s s'}$ and $\sum_jL_{ij}\phi^s_j=\Lambda^s\phi^s_i$ for all $i=1,\dots N$, then
\begin{equation*}
\sum_jM_{ij}\phi^s_j=\frac{1}{\Lambda^s-\alpha c}\phi^s_i
\quad\forall i=1,\dots N\, .
\end{equation*}
In particular, being $\Lambda^1=0$ and $\boldsymbol{\phi}^1 {\propto} (1,\dots,1)^T$, we obtain
\begin{equation}
\sum_jM_{ij}=-\frac{1}{\alpha c}\quad\forall i=1,\dots N\, .
\label{eq:sumMij}
\end{equation}
We can thus rewrite Eq.~\eqref{eq:fhnmodel} as
\begin{equation}
\label{eq:fhnmodel2}
\dot{u}_i = \mu u_i-u_i^3+\frac{\beta}{\alpha}+c\sum_{j}M_{ij}u_j+D_u \sum_j L_{ij}u_j\quad\forall i=1,\dots N\, ,
\end{equation}
which is the {\em reduced FitzHugh-Nagumo} model (for short {\em rFHN} in the following).

{Thus, the matrix $\mathbf{M}$ plays the role of the Green's function.} 
In fluid mechanics, a similar approximation is well known as {the Stokes} approximation~\cite{lamb}, while in {the studies of coupled oscillators,} a similar method {has been used} to derive a model of non-locally coupled oscillator media~\cite{Kuramoto1995,Kuramoto1998}. 
Similar adiabatic elimination has been also used in Refs.~\cite{nicola,siebert} to derive reaction-diffusion systems with nonlocal coupling, including the effect of advection, in continuous media.

\begin{remark}[Multigraph]
Starting from the matrix $\mathbf{M}$ one can define the matrix $\hat{\mathbf{M}}=-\left(\mathbf{M}+\frac{1}{\alpha c}\mathbf{I}\right)$, whose spectrum is given by $\hat{\mu}^s=-\frac{\Lambda^s}{\alpha c (\Lambda^s-\alpha c)}$. Hence $\hat{\mu}^1=0$ and $\hat{\mu}^s<0$ for all $s=2,\dots,N$ and thus  $\hat{\mathbf{M}}$ can be considered as a Laplace matrix of some (possibly weighted) network. This means that once we rewrite the reduced system as
\begin{equation*}
\dot{u}_i = \left(\mu -\frac{1}{\alpha}\right)u_i-u_i^3+\frac{\beta}{\alpha}-c\sum_{j}\hat{M}_{ij}u_j+D_u \sum_j L_{ij}u_j\quad\forall i=1,\dots N\, ,
\end{equation*}
it can be considered as a model where the species $u$ move on a multigraph, i.e. a network where several links can connect two nodes. One network being the original one which {is} encoded by the adjacency matrix $\mathbf{A}$ used to build the Laplace matrix $\mathbf{L}$, {and} the second network is encoded by some adjacency matrix $\hat{\mathbf{A}}$ whose Laplace matrix is $\hat{\mathbf{M}}$. Observe that $\hat{\mathbf{M}}$ can be seen a series of powers of $\mathbf{L}$, indeed
\begin{equation}
\hat{\mathbf{M}}=-\left[({\mathbf{L}}-\alpha c \mathbf{I})^{-1}+\frac{1}{\alpha c} \mathbf{I}\right]=\frac{1}{\alpha c}\sum_{n\geq 1}\frac{(-1)^n}{(\alpha c)^n}\mathbf{L}^n \, ,
\end{equation}
thus the matrix $\hat{\mathbf{A}}$ will encode ``weighted long-range connections'' taken from the first network. Finally, the diffusion on this second network is characterised by ``a negative diffusion coefficient'', $-c$, which tends thus to increase the unbalance of the local densities of $u_i$ instead of decreasing it as done by the term $D_u\mathbf{L}$.
\end{remark}

\begin{remark}[Variational formulation]
Let us observe that, after adiabatic elimination, the remaining equation for activator $u$ can also be written in a variational form when the network and hence $\mathbf{L}$ and $\mathbf{M}$ are symmetric. Defining a potential or 'free energy' function as 
\begin{align}
F(u_1, ..., u_N) = - \sum_i \left( \mu \frac{1}{2} u_i^2 - \frac{1}{4} u_i^4 + \frac{\beta}{\alpha} u_i \right) 
- \frac{1}{2} c \sum_{ij}  u_i M_{ij} u_j
- \frac{1}{2} D_u \sum_{ij}  u_i L_{ij} u_j,
\end{align}
we notice that Eq.~(\ref{eq:fhnmodel2}) can be written as
\begin{align}
\dot{u}_i = - \frac{\partial F(u_1, ..., u_N)}{\partial u_i}
\quad\forall i=1, ..., N,
\end{align}
hence the system evolves towards the direction in which $F$ decreases and the stationary state of the system corresponds to a (local) minimum of $F$.
In Ref.~\cite{Ohta}, spatially continuous version of the above free energy has been obtained and its relation to other models has been discussed.
\end{remark}

\section{Turing patterns}
\label{sec:TP}
The aim of this section is to prove that both systems, the {\em FHN}~\eqref{eq:FHNnet} and the {\em rFHN}~\eqref{eq:fhnmodel2}, exhibit Turing patterns. Before to proceed with our analysis, let us observe that this result for the reduced model is at odd with the general statement that single species reaction-diffusion systems cannot exhibit Turing patterns, both on continuous or networked support. The existence of Turing patterns relies thus on the multigraph nature of the support and the long-range connections hereby present.

\subsection{Turing {instability} in the {\em FHN} model}
\label{ssec:TPFHN}

Turing instability emerges under the assumption of the existence of an equilibrium solution, stable with respect to homogeneous perturbations, or equivalently in absence of diffusion, $D_u=D_v=0$, that loses its stability for heterogeneous perturbations once diffusion is in action, $D_u>0$ and $D_v>0$. In the following, we will assume $\beta=0$ and we will be interested in the point $(u_i,v_i)=(0,0)$ for all $i=1,\dots, N$, which is clearly an equilibrium for Eq.~\eqref{eq:FHNnet}. 

{The linear stability can be performed by following the standard procedure, that is, linearising the model~\eqref{eq:FHNnet} around this equilibrium, decomposing the perturbations using the eigenbasis of $\mathbf{L}$, and calculating the dispersion relation, i.e. the linear growth rate  $\lambda_s = \lambda(\Lambda^s)$ of the eigenmode $s$ as a function of the Laplacian eigenvalue $\Lambda^s$.
The linear growth rate is explicitly given by (see Appendix~\ref{sec:linearstability2com} for  the derivation)
\begin{align}
\label{eq:reldispFHNmain}
\lambda_s(\Lambda^s)
&= \frac{1}{2}\Big[\mu -\alpha \gamma+\Lambda^s (D_u+D_v)
\\
&\pm\sqrt{\left[\mu -\alpha \gamma+\Lambda^s (D_u+D_v)\right]^2-4\left[\gamma(1-\alpha \mu)+\Lambda^s(\mu D_v-\gamma\alpha D_u)+D_uD_v(\Lambda^s)^2\right]}\Big]\notag\, .
\end{align}
}
A straightforward computation provides the following expressions for the {roots of $\lambda_s(\Lambda^s)$:}
\begin{eqnarray}
\Lambda_{-} = -\left(\frac{\mu}{2D_u}-\frac{\alpha\gamma}{2D_v}\right)-\frac{1}{2}\sqrt{\left(\frac{\mu}{D_u}-\frac{\alpha\gamma}{D_v}\right)^2-\frac{4\gamma(1-\mu\alpha)}{D_uD_v}},
\cr
\Lambda_{+} = -\left(\frac{\mu}{2D_u}-\frac{\alpha\gamma}{2D_v}\right)+\frac{1}{2}\sqrt{\left(\frac{\mu}{D_u}-\frac{\alpha\gamma}{D_v}\right)^2-\frac{4\gamma(1-\mu\alpha)}{D_uD_v}}\, ,
\label{eq:rootreldispFHNmain}
\end{eqnarray}
under the conditions $\left(D_v{\mu}-D_u{\alpha\gamma}\right)^2-{4\gamma(1-\mu\alpha)}{D_uD_v}>0$ (for the reality of the roots) and ${\mu}{D_v}-{\alpha\gamma}{D_v}>0$ (for the negativity of the roots).
{Turing instability can occur if there exists $\Lambda^s$ such that {$\Lambda_{-}<\Lambda^s<\Lambda_{+}$}, for some $s=2,\dots, N$.}

\begin{figure}[ht]
\centering
\includegraphics[scale=0.5]{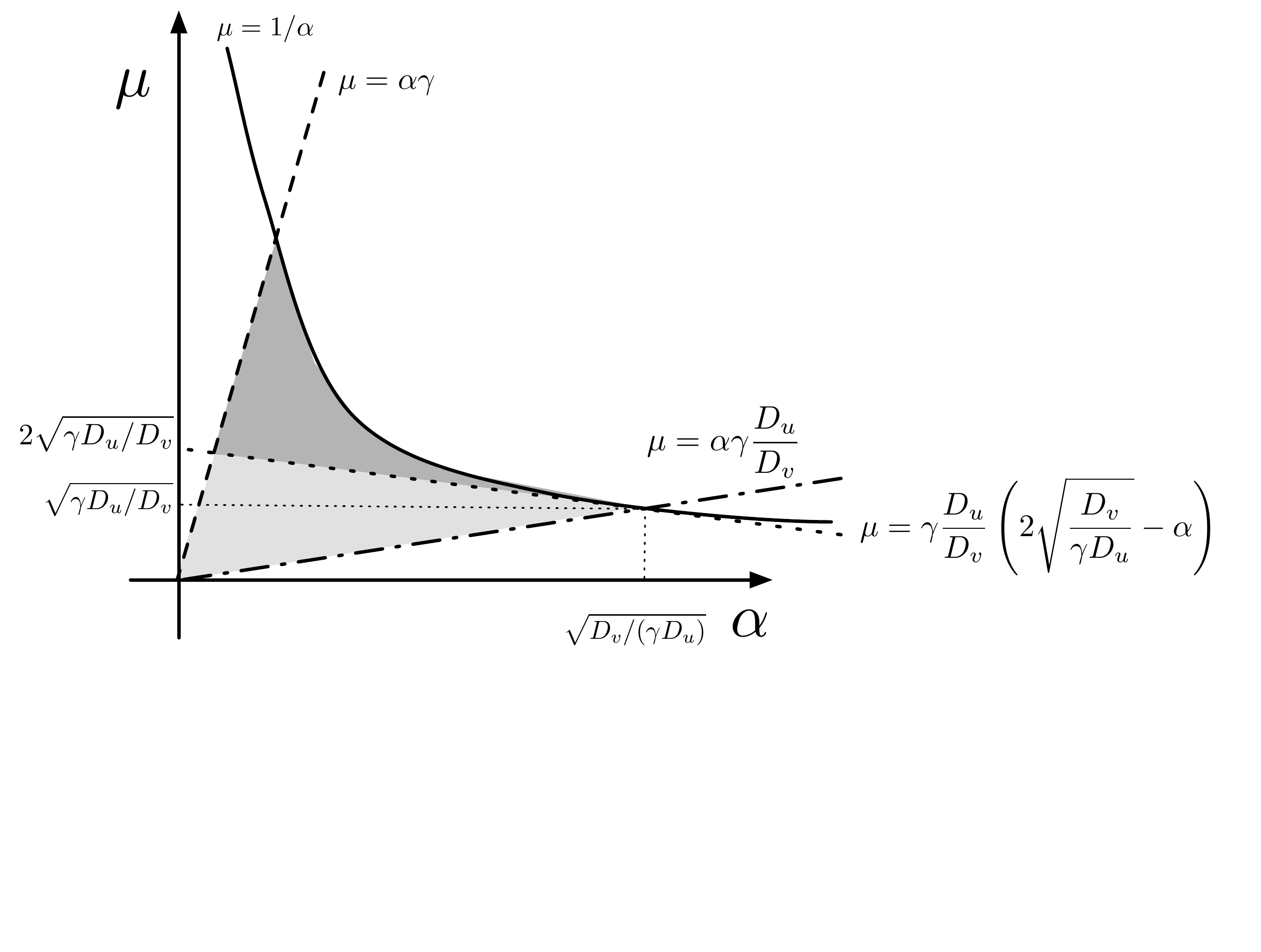}
\vspace{-2.5cm}
\caption{\textbf{Parameter region for the onset of Turing {instability} for the {\em FHN} model.} The grey region (light and dark ones) is defined by $\mu\alpha  <1$ (solid line), $\mu < \alpha \gamma$ (dashed line) and corresponds to parameters associated with a stable homogeneous equilibrium. This region contains (dark grey) the subregion of Turing instability, delimited by $\mu\alpha  <1$, $\mu < \alpha \gamma$ and $\mu>-\alpha \gamma D_u/D_v+2\sqrt{\gamma D_u/D_v}$ (dotted line).}
\label{fig:mualpgaFHN}
\end{figure}

Summarising, the conditions for the emergence of Turing patterns are given by (see Fig.~\ref{fig:mualpgaFHN}):
\begin{itemize}
\label{item:TuringcondFHN}
\item[a)] $\mu\alpha  <1$;
\item[b)] $\mu < \alpha \gamma$;
\item[c)] $\mu>\alpha \gamma D_u/D_v$;
\item[d)] $\mu>-\alpha \gamma D_u/D_v+2\sqrt{\gamma D_u/D_v}$.
\end{itemize}
The {conditions} a) and b) ensure the stability of $(u_i,v_i)=(0,0)$ under homogeneous perturbations, while c) and d) the fact that it loses its stability under heterogeneous perturbations. It is worth remarking that the hyperbola, $\mu\alpha  =1$, and the two straight lines, $\mu=\alpha \gamma D_u/D_v$ and $\mu=-\alpha \gamma D_u/D_v+2\sqrt{\gamma D_u/D_v}$,  all meet at the same point, ($\alpha,\mu)=(\sqrt{D_v/(\gamma D_u)},\sqrt{\gamma D_u/Dv})$, hence condition d) is stronger than condition c) in the region $\mu\alpha  <1$. Let us observe that, if $D_u>D_v$, the conditions b) and c) are not compatible and thus the Turing region is empty.

\subsection{Turing {instability} in the {\em rFHN} model}
\label{ssec:TPrFHN}

Let us now consider the {\em rFHN} model Eq.~\eqref{eq:fhnmodel2}, still under the assumption $\beta=0$ and look for a stationary homogeneous solution {satisfying}
\begin{equation*}
0 = \mu u-u^3-\frac{u}{\alpha}=\left(\mu-\frac{1}{\alpha}\right)u-u^3\, ,
\end{equation*}
where we used Eq.~\eqref{eq:sumMij} to write $c\sum_j M_{ij}u=-u/\alpha$. The function $\hat{u}=0$ is still {an equilibrium} solution and it is stable if and only if $\mu-\frac{1}{\alpha}<0$. {There is a bifurcation at $\mu=1/\alpha$ at which this solution becomes unstable and two new stable solutions are created, $\tilde{u}_{\pm}=\pm\sqrt{\mu-1/\alpha}$ if $\mu-1/\alpha>0$.} {We assume $\mu-\frac{1}{\alpha}<0$ hereafter.}

Let us study the {stability of the equilibrium $\hat{u}_i = 0$ for all $i=1, ..., N$} under heterogeneous perturbations, {namely we will set $u_i=\hat{u}_i+\delta u_i$}. {We linearise} Eq.~\eqref{eq:fhnmodel2} around $\hat{u}=0$ to obtain
\begin{equation}
\label{eq:fhnmodellin}
\frac{d{\delta u}_i }{dt}= \mu \delta u_i+c\sum_{j}M_{ij}\delta u_j+D_u \sum_j L_{ij}\delta u_j\quad\forall i=1,\dots N\, ,
\end{equation}
then we again decompose the perturbation on the eigenbasis of the Laplace matrix $\mathbf{L}$ and make the following ansatz for the time evolution of the perturbation, $\delta u_i=\sum_s \rho_s \phi_i^s e^{\lambda^{(r)}_st}$. Inserting the latter in Eq.~\eqref{eq:fhnmodellin} we obtain
\begin{equation}
\label{eq:fhnmodellinLap}
\sum_s \rho_s \lambda_s\phi_i^s e^{\lambda^{(r)}_st} = \mu \sum_s \rho_s \phi_i^s e^{\lambda^{(r)}_st}+c\sum_s \rho_s \frac{1}{\Lambda^s-\alpha c}\phi_i^s e^{\lambda^{(r)}_st}+D_u \sum_s \rho_s \Lambda^s\phi_i^s e^{\lambda^{(r)}_st}\quad\forall i=1,\dots N\, ,
\end{equation}
{and} using the orthogonality of $\boldsymbol{\phi}^s$, we {obtain the} dispersion {relation} for the {\em rFHN} model:
\begin{equation}
\label{eq:reldisp}
\lambda^{(r)}_s(\Lambda^s) = \mu+\frac{c}{\Lambda^s-\alpha c}+D_u \Lambda^s\quad\forall s=1,\dots N\, .
\end{equation}
The dispersion relation is real for all $s$, hence its sign will determine the stability of $\hat{u}=0$ subjected to heterogeneous perturbations. A straightforward computation (see {Appendix}~\ref{sec:reldisprFHN}) allows to draw our conclusions. We can have a positive {growth rate} $\lambda^{(r)}_s(\Lambda^s)$ for some $s$ if and only if the following three conditions are met:
\begin{itemize}
\label{item:Turingcond}
\item[a)] $\mu <\frac{1}{\alpha}$ for the stability of $\hat{u}=0$ under homogeneous perturbations;
\item[b)] $\alpha -1/\sqrt{c D_u}<0$;
\item[c)] $\mu>D_u c\left(\frac{2}{\sqrt{D_u c}}-\alpha\right)$ for the positivity of $\lambda^{(r)}_s(x)$ for some negative $x$.
\end{itemize}

Once the above conditions are met, Turing {instability occurs} if there exists $\Lambda^s$ such that {$\Lambda^{(r)}_{-}<\Lambda^s<\Lambda^{(r)}_{+}$}, for some $s=2,\dots, N$, where
\begin{eqnarray}
\label{eq:rootreldisprFHN}
\Lambda^{(r)}_{-}&=&-\frac{1}{2}\left(\frac{\mu}{D_u}-\alpha c \right)-\frac{1}{2}\sqrt{\left(\frac{\mu}{D_u}-\alpha c D_u\right)^2-4\frac{c(1-\alpha \mu)}{D_u}},
\\
\Lambda^{(r)}_{+}&=&-\frac{1}{2}\left(\frac{\mu}{D_u}-\alpha c \right)+\frac{1}{2}\sqrt{\left(\frac{\mu}{D_u}-\alpha c D_u\right)^2-4\frac{c(1-\alpha \mu)}{D_u}}\, .
\end{eqnarray}
One can easily realise that $\Lambda^{(r)}_{-}<\Lambda^{(r)}_{+}<0$.

\begin{figure}[ht]
\centering
\includegraphics[scale=0.5]{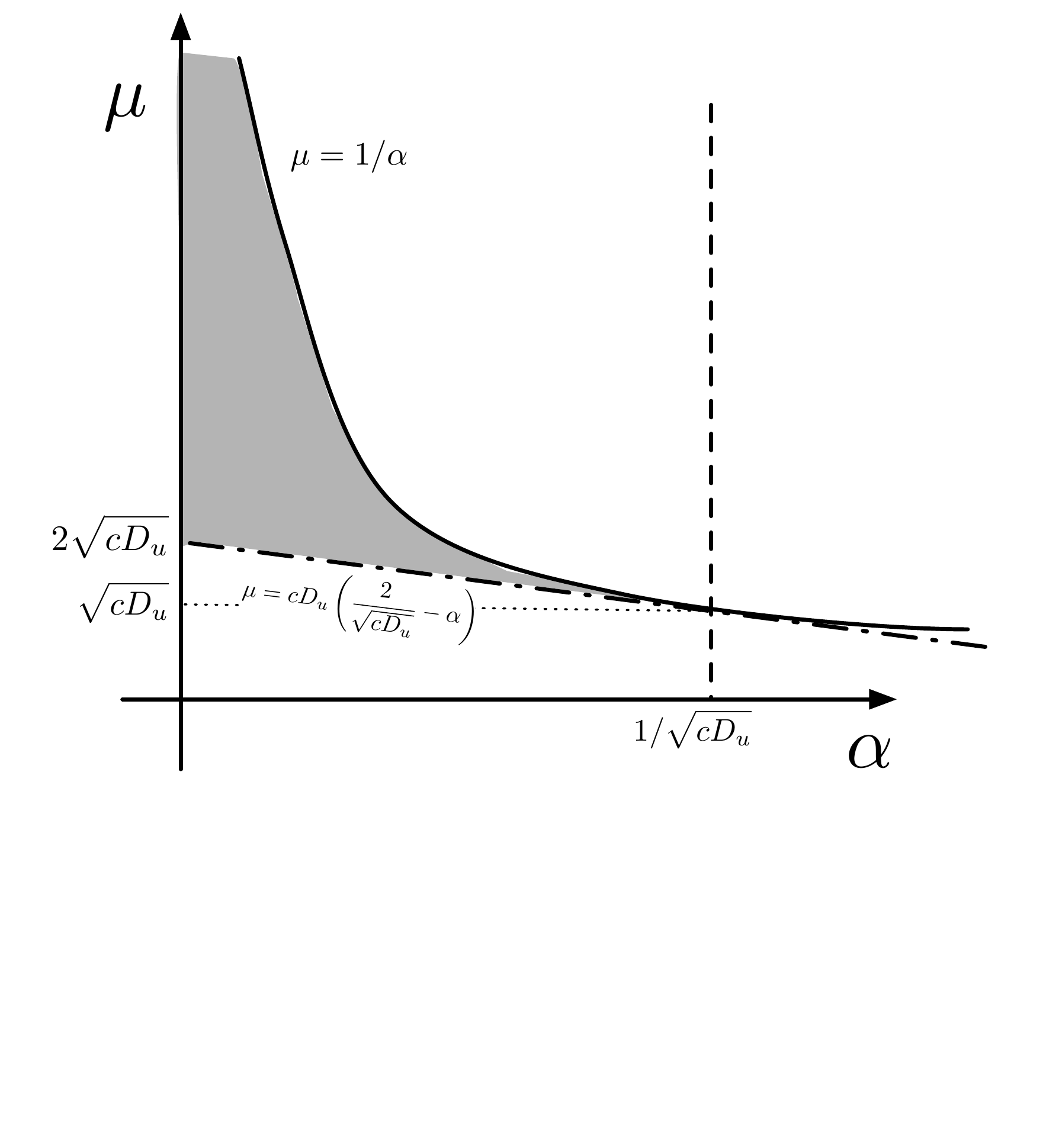}
\vspace{-1.5cm}
\caption{\textbf{Parameter region for the onset of Turing {instability} {for the {\em rFHN} model}.} The grey region is is defined by $D_u c\left(\frac{2}{\sqrt{D_u c}}-\alpha\right) < \mu < 1/\alpha$ for $0<\alpha<1/\sqrt{cD_u}$.}
\label{fig:mualpga}
\end{figure}

Let us now describe in the parameters plane $(\mu,\alpha)$ the domain for which the previous conditions are satisfied for fixed $c$ and $D_u$ (see Fig.~\ref{fig:mualpga}). Condition b) implies that only $\alpha$ in the strip $0<\alpha<1/\sqrt{c D_u}$ should be considered (vertical dashed line in Fig.~\ref{fig:mualpga}); condition a) tells that $\mu$ should lie below the hyperbola $\mu=1/\alpha$ (solid line in Fig.~\ref{fig:mualpga}), finally condition c) means that $\mu$ should lie above the straight line $\mu=D_u c\left(\frac{2}{\sqrt{D_u c}}-\alpha\right)$ (dot-dashed line in Fig.~\ref{fig:mualpga}). Observe that the latter line is tangent to the hyperbola at the point $(1/\sqrt{cD_u},\sqrt{cD_u})$, hence the straight line completely lies below the hyperbola. In conclusion, the parameters set for which the system Eq.~\eqref{eq:fhnmodel} exhibits Turing {instability} is given by the grey region in Fig.~\ref{fig:mualpga}.

Let us observe that, even if we vary $cD_u$, the instability region keeps qualitatively the same shape, the main change being that if $cD_u\rightarrow 0$ then the straight line $\mu=D_u c\left(\frac{2}{\sqrt{D_u c}}-\alpha\right)$ tends to the horizontal $\alpha$-axis and the vertical line $\alpha=1/\sqrt{cD_u}$ moves toward infinity. Stated differently, in the limit $cD_u\rightarrow 0$, the parameter region for which the system has Turing pattern increases and tends to $0<\mu<1/\alpha$ for $\alpha>0$. On the opposite, if $cD_u\rightarrow \infty$, the Turing region shrinks, because the point $\alpha=1/\sqrt{cD_u}$  gets closer and closer to $0$ and the straight line $\mu=D_u c\left(\frac{2}{\sqrt{D_u c}}-\alpha\right)$  gets steeper and steeper.

\begin{figure}[ht]
\centering
\includegraphics[scale=0.3]{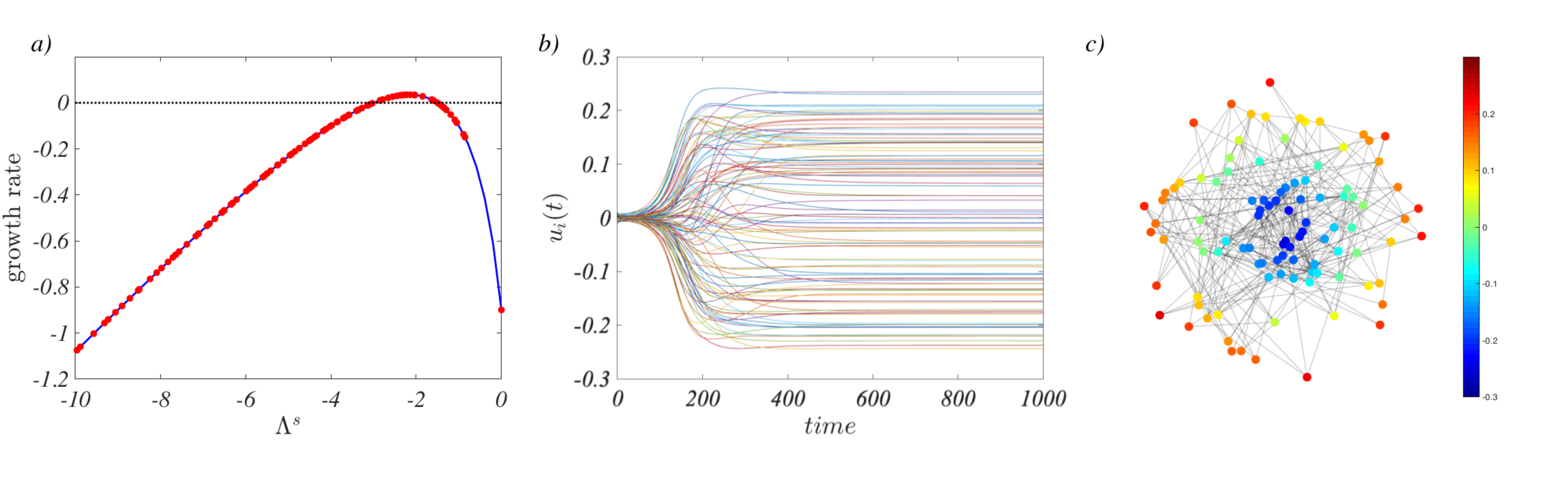}
\caption{\textbf{Turing patterns in the reduced {{\em rFHN} model}.} {(a) Dispersion relation} for the reduced system defined on the continuous support (blue line) and on the networked system (red dots); the network is an Erd\H{o}s-R\'enyi network made by $N=100$ nodes and probability for a link $p=0.05$. {(b)} Time evolution of $u_i(t)$ for each node {($i=1, ..., 100$)}.  {(c) Turing patterns on the network, nodes are coloured according to the asymptotic value of $u_i(t)$, dark blue are associated with negative values while dark red to positive ones}. The model parameters have been set to: $\mu=1.1$, $\beta=0$, $\alpha=0.5$, $c=2$ and $D_u=0.2$.}
\label{fig:utpatt}
\end{figure}

In Fig.~\ref{fig:utpatt} (a), we report the dispersion {relation} for a set of parameters for which Turing {instability does occur}; the resulting Turing patterns are shown on the panels (b, c), $u_i(t)$ versus $t$, using an Erd\H{o}s-R\'enyi network made by $N=100$ nodes and probability for a link $p=0.05$, as underlying support.

\subsection{Comparison of the two models}
\label{ssec:compare}

The aim of this subsection is to compare the results obtained so far for the two models. Let us fist start with an eyeball comparison of the Turing domains {shown in} Figs.~\ref{fig:mualpgaFHN} and~\ref{fig:mualpga}; replacing $\gamma=cD_v$ and letting $\gamma\rightarrow \infty$ {in Fig.~\ref{fig:mualpgaFHN}}, and so does $D_v$, we realise that the straight line $\mu=\alpha \gamma$ becomes steeper and steeper to eventually merge with the $\mu$-axis, and thus the Turing domain for the {\em FHN} model ``converges'' to the one for the {\em rFHN}. More quantitatively, we claim that the dispersion relation for the {\em FHN} model behaves smoothly with respect to the adiabatic elimination, namely
\begin{equation}
\label{eq:limreldisp}
\lim_{\substack{D_v\rightarrow \infty \\ c=\gamma/D_v\in \mathbb{R}}}\lambda_s(x)=\lambda^{(r)}_s(x)\quad \forall x\, .
\end{equation}

To prove this claim, let us perform some algebraic manipulation on the formula Eq.~\eqref{eq:reldispFHNmain} for the $\lambda_s(\Lambda^s)$ and rewrite it as follows (we are considering only the root with the plus sign):
\begin{equation}
\label{eq:reldispFHNbis}
\lambda_s(\Lambda^s)=\frac{1}{2}\left[\mu -\alpha \gamma+\Lambda^s (D_u+D_v)+\sqrt{-4\gamma+\left[\mu+\alpha \gamma+(D_u-D_v)\Lambda^s\right]^2}\right]\, .
\end{equation}
Let us now substitute $\gamma=cD_v$ and consider $D_v\gg 1$.
{After some transformations given in Appendix~\ref{sec:reldispcomparison}, it can be shown that} 
\begin{eqnarray}
\label{eq:reldispFHNbis2}
\lambda_s(\Lambda^s)
&=&\mu +\Lambda^s D_u+\frac{c}{\Lambda^s-\alpha c}+\mathcal{O}(1/D_v)=\lambda^{(r)}_s +\mathcal{O}(1/D_v)\, ,
\end{eqnarray}
from which the claim follows by taking the limit $D_v\rightarrow \infty$.

\begin{figure}[h]
\centering
\includegraphics[scale=0.25]{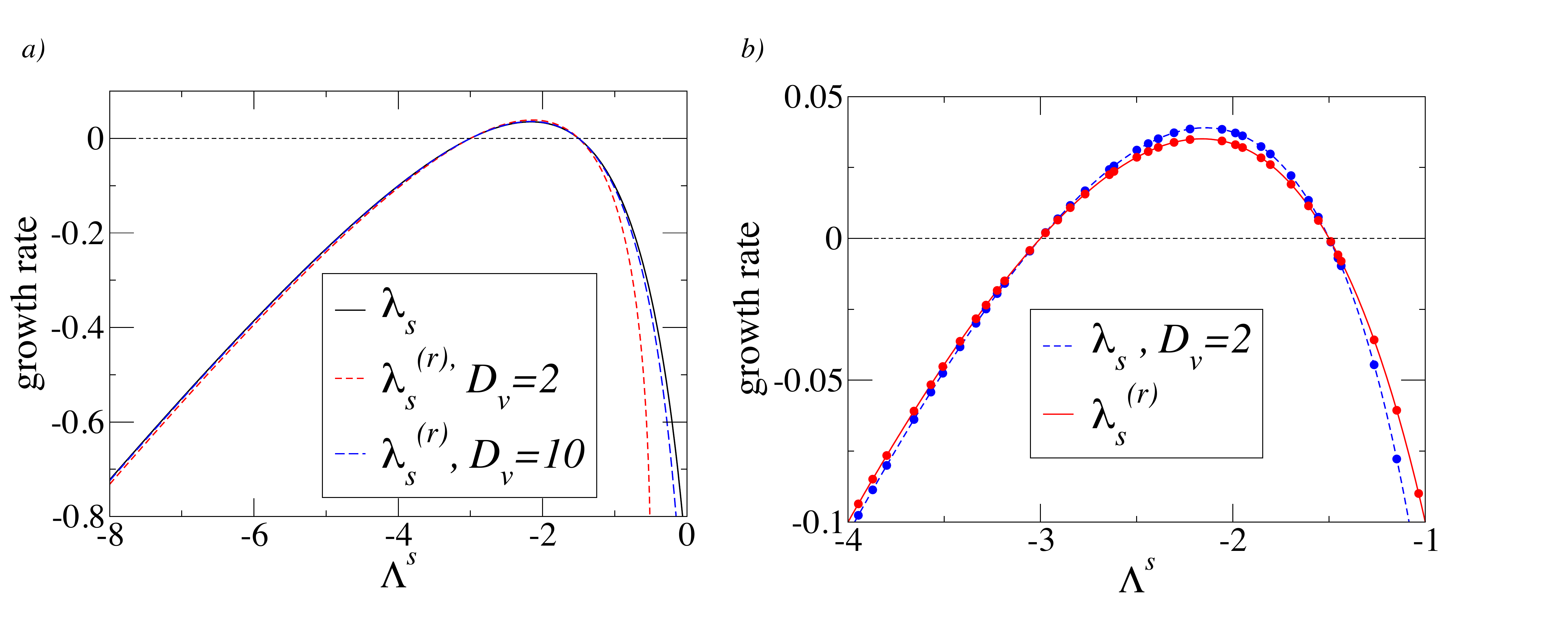}
\caption{\textbf{Comparison of $\lambda_s$ and $\lambda_s^{(r)}$}. (a) Two curves of $\lambda_s$ corresponding to $D_v = 2$ and $Dv = 10$ are compared with $\lambda_s^{(r)}$ of the reduced model. (b) Close-up of the region near the peak for $Dv=2$ and $\lambda_s^{(r)}$. Dots correspond to the actual Laplacian eigenvalues of the Erd\H{o}s-R\'enyi network of $N=100$ and $p=0.05$. The other parameters are
$\mu = 1.1$, $\beta = 0$, $\alpha = 0.5$, $c = 2.0$, and $D_u = 0.2$.}
\label{fig:fig-dispersion}
\end{figure}

Secondly, we can prove a more interesting result, {that is,} both $\lambda_s$ and $\lambda^{(r)}_s$ possess the same roots, hence both models exhibit {(or do not exhibit)} Turing patterns for the same set of parameters and networks. To prove this statement,
%
{we substitute $\gamma=cD_v$ into Eq.~\eqref{eq:rootreldispFHNmain} and make some simplifications to obtain}
\begin{eqnarray*}
\Lambda_{-} = -\left(\frac{\mu}{2D_u}-\frac{\alpha c}{2}\right)-\frac{1}{2}\sqrt{\left(\frac{\mu}{D_u}-\alpha c\right)^2-\frac{4c(1-\mu\alpha)}{D_u}},
\cr
\Lambda_{+} = -\left(\frac{\mu}{2D_u}-\frac{\alpha c }{2}\right)+\frac{1}{2}\sqrt{\left(\frac{\mu}{D_u}-\alpha c\right)^2-\frac{4c(1-\mu\alpha)}{D_u}}\, ;
\end{eqnarray*}
by direct comparison, we can then conclude that they coincide with the equations for $\Lambda_{-}^{(r))}$ and $\Lambda_{+}^{(r)}$ {of the reduced model} given by Eq.~\eqref{eq:rootreldisprFHN} (see {the panel (a) of Fig.~\ref{fig:fig-dispersion}}).

\begin{figure}[h]
\centering
\includegraphics[scale=0.25]{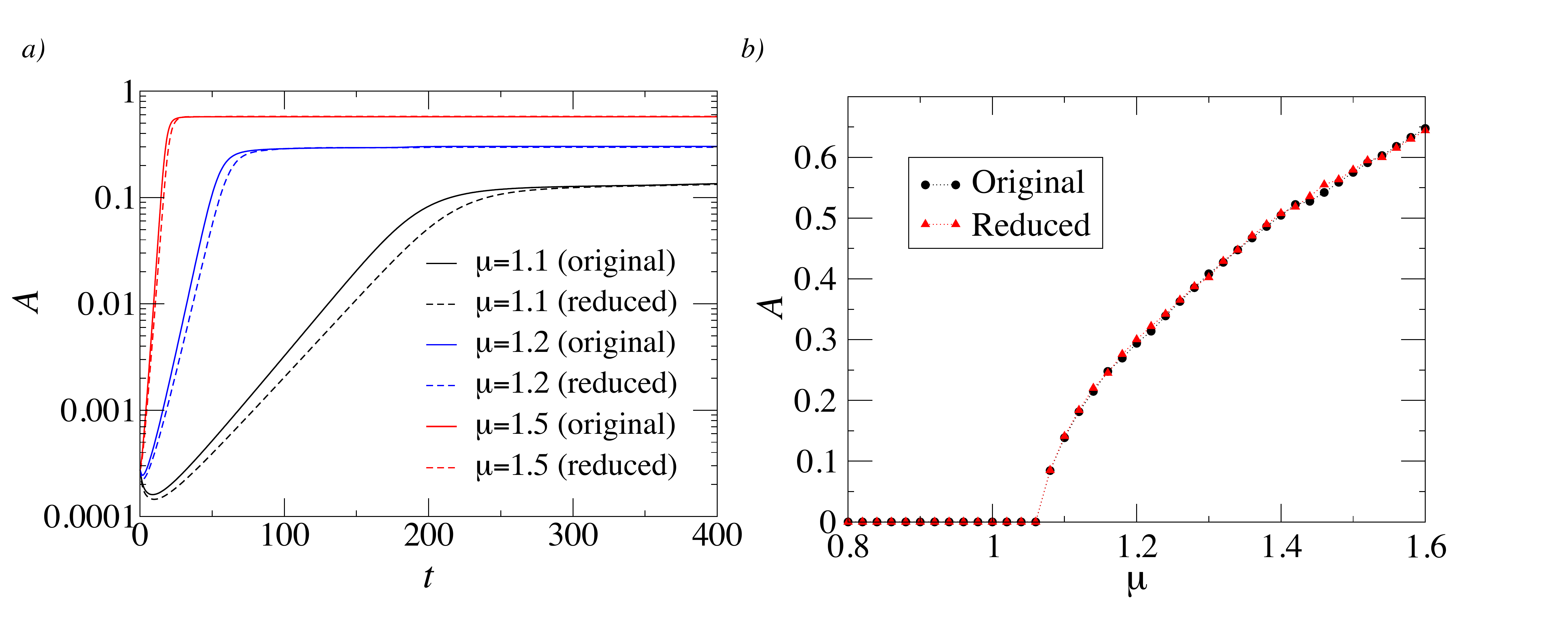}
\caption{\textbf{Amplitude of the Turing pattern}. (a) Evolution of the amplitude of the Turing pattern for $\mu=1.1$, $\mu=1.2$, and $\mu=1.5$. (b) Stationary amplitude of the Turing pattern versus $\mu$. In both figures, results for the original and reduced models are compared. The diffusion constant for $v$ is $D_v = 2$ in both figures. The network and other parameters are the same as in Fig.~\ref{fig:fig-dispersion}.}
\label{fig:fig-amplitude}
\end{figure}

The dispersion relations of the two models differ in the value of the maximum and its location (see {the panel (b) of Fig.~\ref{fig:fig-dispersion}}); a straightforward computation for the {\em FHN} model {gives}
\begin{align}
\label{eq:peakraldispFHN}
\Lambda_{max} &=-\frac{D_v\left(\alpha c-\sqrt{c/D_u}\right)-\sqrt{cD_u}+\mu}{D_v-D_u},
\cr
\lambda_{max} &:=\lambda(\Lambda_{max})=\frac{D_v}{D_v-D_u}\left(\alpha c D_u-2\sqrt{cD_u}+\mu\right)\, ,
\color{black}
\end{align}
where we assumed $\gamma=cD_v$ to compare with {those} for the {\em rFHN} {model,}
\begin{align}
\label{eq:peakraldisprFHN}
\Lambda^{(r)}_{max} &=\alpha c-\sqrt{c/D_u}
\cr
\lambda^{(r)}_{max} &:=\lambda^{(r)}(\Lambda^{(r)}_{max})=\alpha c D_u-2\sqrt{cD_u}+\mu\, .
\end{align}
As a consequence, a faster growth for short times of the orbits of the {\em FHN} {is expected, as can be seen in numerical simulations} (see panel (a) of Fig.~\ref{fig:fig-amplitude}).

In Fig.~\ref{fig:fig-dispersion}, we compare the dispersion curves, $\lambda_s$ of the original model and $\lambda_s^{(r)}$ of the reduced one. For the original model, we consider two cases, $D_v = 2$ and $D_v = 20$. As discussed above, all dispersion curves intersect with the horizontal axis at the the same roots. For $D_v = 20$, $\lambda_s(x)$ and $\lambda_s^{(r)}(x)$ are almost indistinguishable except for the region where $x$ is very close to $0$. For $D_v = 2$, we can also appreciate a tiny difference near the peaks of $\lambda_s(x)$ and $\lambda_s^{(r)}(x)$ as shown in the panel (b) of Fig.~\ref{fig:fig-dispersion}, where close-ups of the two curves are plotted with the actual Laplacian eigenvalues of the Erd\H{o}s-R\'enyi network made by $N=100$ nodes and $p=0.05$ as probability for the existence of a link. It can be seen that the maximum growth rate of the original model is slightly larger than the one of the reduced model as proved above.

This observation translates into a faster growth, for short times, of the {Turing pattern} for the {\em FHN} model with respect to the ones of the {\em rFHN}. In Fig.~\ref{fig:fig-amplitude}, we show the amplitude of the Turing pattern defined as
\begin{align}
A = \sqrt{ \frac{1}{N} \sum_{i=1}^N \left( u_i - \langle u \rangle \right)^2  }, \quad \langle u \rangle = \frac{1}{N} \sum_{i=1}^N u_i,
\end{align}
for $D_v = 2$. In the left panel, time evolution of $A$ is plotted as a function of $t$ for $\mu=1.1, 1.2$, and $1.5$. The initial condition is $(u_i, v_i) = (\xi_i, 0)$ for $i=1, ..., N$, where $\xi_i$ is a random number independently chosen from a uniform distribution on $[0, 0.001]$. Slight discrepancy can be seen between the original and reduced models in the initial transient, reflecting that $\lambda_{max} > \lambda^{(r)}_{max}$, but eventually the curves for the original and reduced models converge to approximately the same stationary values.
In the right panel, the stationary value of $A$, computed after a sufficiently large time interval {after the initial transient}, is plotted as a function of $\mu$. The results for the original and reduced models are almost equal except for small fluctuations.

Let us conclude this analysis by observing that the above conclusion is not a generic one but it relies to a peculiar {property} of the FitzHugh-Nagumo model. In Appendix~\ref{sec:genmodel} we briefly study under which assumption a generic reaction-diffusion system defined on a network will exhibit the same behaviour {as} the FitzHugh-Nagumo {model}.

\section{Conclusion}
\label{sec:concl}

In this paper, we considered the FitzHugh-Nagumo model defined on a symmetric network, the activator and inhibitor species react on the nodes and spread (bidirectionally) across the available links. We then applied the adiabatic elimination to this model; the resulting system describes a single species (of activator type) reacting on each node and diffusing across the links of a multigraph, where long-range {nonlocal} interactions {derive from} the adiabatic elimination the original activator-inhibitor model. In this framework, we proved that Turing patterns can develop. We analytically describe the set of parameters associated with the onset of the Turing instability and we compare it with the similar domain for the original model. We proved that Turing patterns can emerge in the reduced model if and only if the original model does, in this way the adiabatic elimination process can be used to simplify the analysis of the original model.

The adiabatic elimination can be cast in the more general framework of slow-fast systems; we have shown that the above property of the FitzHugh-Nagumo model is not generic and we provided the conditions a system must satisfy to exhibit patterns for the same {set of} parameters as the original model.

Long-range nonlocal interactions have often been considered in mathematical models of pattern formation in continuous media, which arise from elimination of fast variables or assumed, ab initio, as models of real biological processes ~\cite{Murray,Kondo2,bresslof}. As considered in this paper, such nonlocal interactions can also be introduced into networked dynamical systems in the form of multigraphs. Such multigraph natures could lead to novel self-organisation of dynamical processes on networks.

Based on the above and the fact that our result can be straightforwardly generalised as to include directed networks, we are confident that the work hereby proposed can open the way to relevant applications in many diverse research domains {on dynamical processes on networks}.

\acknowledgments
We thank M. Asllani, J. Petit, R. Muolo, and D. Fanelli for useful discussions. H.N. thanks JSPS KAKENHI, Grant Nos. JP16K13847, JP17H03279, 18K03471, and JP18H03287 for financial support. 

\appendix

\section{Linear stability analysis of the original FitzHugh-Nagumo model}
\label{sec:linearstability2com}

{In this section, the details of the linear stability analysis of the original FitzHugh-Nagumo model are given.
We first linearise the model about the equilibrium, $(u_i,v_i)=(0,0)$ for all $i=1,\dots, N$.}
Let us thus consider a small perturbation about this equilibrium, $(\delta u_i(t),\delta v_i(t))$, and then linearise the model~\eqref{eq:FHNnet}:
\begin{equation}
\begin{cases}
\dot{\delta u}_i(t) &=\mu \delta u_i- \delta v_i+D_u \sum_j L_{ij} \delta u_j\\
\dot{\delta v}_i(t) &=\gamma (\delta u_i-\alpha \delta v_i)+D_v  \sum_j L_{ij}\delta v_j\, .
\end{cases}
\label{eq:linFHNnet}
\end{equation}
To infer the stability of this equilibrium, let us decompose the solutions $(\delta u_i(t),\delta v_i(t))$ using the eigenbasis of $\mathbf{L}${,}
\begin{equation*}
\delta u_i (t)= \sum_s \xi_s \phi^s_i e^{\lambda_s t} \ \text{ and } \ \delta v_i (t)= \sum_s \eta_s \phi^s_i e^{\lambda_s t}\, .
\end{equation*}
We will soon see that this approach allows to ``decompose'' the study of the stability of a $2N\times 2N$ matrix into the study of $N$ simpler $2\times 2$ systems from which we could obtain a full analytical comprehension of the model. Inserting hence this ansatz into Eq.~\eqref{eq:linFHNnet}, we obtain 
\begin{equation*}
\begin{cases}
\sum_s \xi_s \lambda_s\phi^s_i e^{\lambda_s t} &= \sum_s (\mu \xi_s - \eta_s) \phi^s_i e^{\lambda_s t} +D_u\sum_s\xi_s\Lambda^s \phi^s_i e^{\lambda_s t}\\
\sum_s \eta_s \lambda_s\phi^s_i e^{\lambda_s t}  &=\sum_s \gamma ( \xi_i-\alpha \eta_i)\phi^s_i e^{\lambda_s t}+D_v \sum_s\eta_s\Lambda^s \phi^s_i e^{\lambda_s t}\, ,
\end{cases}
\end{equation*}
and employing the orthonormality of the Laplace eigenbasis, we obtain for all $s$:
\begin{equation*}
\begin{cases}
\xi_s \lambda_s&= (\mu \xi_s - \eta_s)  +D_u \xi_s\Lambda^s\\
\eta_s \lambda_s &=\gamma (\xi_i-\alpha \eta_i)+D_v \eta_s\Lambda^s \, .
\end{cases}
\end{equation*}
Introducing the Jacobian of the reaction part of the system, $\mathbf{J}_0$, and $\mathbf{D}$, the $2\times 2$ diagonal matrix whose entries are the diffusion coefficients $D_u$ and $D_v$, we can rewrite the previous system in matrix form
\begin{equation*}
\lambda_s \binom{\xi_s}{\eta_s} = \left[\left(\begin{matrix}\mu & -1\\ \gamma & -\alpha \gamma\end{matrix}\right)+\Lambda^s  \left(\begin{matrix}D_u & 0\\ 0 & D_v\end{matrix}\right)\right]\binom{\xi_s}{\eta_s}=\left(\mathbf{J}_0+\Lambda^s \mathbf{D}\right)\binom{\xi_s}{\eta_s}=:\mathbf{J}_s\binom{\xi_s}{\eta_s}\, .
\end{equation*}
We can hence realise that a non-trivial solution for the previous equation can be found {by} imposing $\lambda_s$ to satisfy the following second order characteristic equation:
\begin{align*}
(\lambda_s)^2-\lambda_s \mathrm{tr}(\mathbf{J}_s)+\det (\mathbf{J}_s) 
=& (\lambda_s)^2-\lambda_s\left(\mu-\gamma \alpha+\Lambda^s(D_u+D_v)\right)
\cr
+&
\gamma(1-\alpha \mu)+\Lambda^s(\mu D_v-\gamma\alpha D_u)+D_uD_v(\Lambda^s)^2=0\, ,
\end{align*}
whose solutions are given by
\begin{align}
\label{eq:reldispFHN}
\lambda_s =& \frac{1}{2}\left[\mathrm{tr}(\mathbf{J}_s)\pm \sqrt{\left[\mathrm{tr}(\mathbf{J}_s)\right]^2-4\det (\mathbf{J}_s)}\right]\\
=& \frac{1}{2}\Big[\mu -\alpha \gamma+\Lambda^s (D_u+D_v)
\cr
&\pm\sqrt{\left[\mu -\alpha \gamma+\Lambda^s (D_u+D_v)\right]^2-4\left[\gamma(1-\alpha \mu)+\Lambda^s(\mu D_v-\gamma\alpha D_u)+D_uD_v(\Lambda^s)^2\right]}\Big]\notag\, .
\end{align}

The {linear growth rate} $\lambda_s$, which can be complex, depends on the eigenvalues of the Laplace matrix $\Lambda^s$, {and} we will thus write $\lambda_s=\lambda(\Lambda^s)$. The real parts of the $\lambda_s$ determine the fate of the perturbation; if they are all negative for all $s$, then the perturbation fades away and the equilibrium is stable.
On the other hand, the existence of $s>1$ for which $\Re \lambda_s >0$ implies the emergence of the Turing instability provided the homogeneous equilibrium is stable once $D_u=D_v=0$. 
{The stability of the homogeneous equilibrium} can be assessed {by} considering $s=1$. {Recalling} that $\Lambda^1=0$, {we have}
\begin{equation}
\label{eq:reldisphomog}
\lambda_1=\frac{1}{2}\left[\mathrm{tr}(\mathbf{J}_1)\pm \sqrt{\left[\mathrm{tr}(\mathbf{J}_1)\right]^2-4\det (\mathbf{J}_1)}\right]=\frac{1}{2}\left[\mu -\alpha \gamma\pm\sqrt{\left[\mu -\alpha \gamma\right]^2-4\gamma(1-\alpha \mu)}\right]\, ,
\end{equation}
and thus the homogeneous solution $(u_i,v_i)=(0,0)$ is stable if and only if
\begin{equation}
\mathrm{tr}(\mathbf{J}_1)=\mu-\alpha \gamma<0 \text{ and }\det (\mathbf{J}_1)=\gamma(1-\alpha\mu)>0\, .
\label{eq:stabhomog}
\end{equation}

Turing patterns {can} emerge if there exists $s>1$ for which $\Re \lambda_s>0$. Looking at Eq.~\eqref{eq:reldispFHN} and observing that $\mathrm{tr}(\mathbf{J}_s)=\mathrm{tr}(\mathbf{J}_1)+\Lambda^s(D_u+D_v)<0$, we can conclude that there exists a negative  $\Lambda^s$ for which $\Re\lambda(\Lambda^s)>0$ if and only if $\det (\mathbf{J}_s)<0$. For the same reason, we can also conclude that, among the two roots, the one with the plus sign has the largest real part. By definition of $\mathbf{J}_s$, we obtain
\begin{equation*}
\det (\mathbf{J}_s)=\gamma(1-\alpha \mu)+\Lambda^s(\mu D_v-\gamma\alpha D_u)+D_uD_v(\Lambda^s)^2\, ,
\end{equation*}
and thus it assumes negatives values, for some $\Lambda^s<0$, if and only if there are two real negative roots. 

\section{{Dispersion relation} for the reduced FitzHugh-Nagumo model}
\label{sec:reldisprFHN}

The aim of this section is to study the dispersion relation for the reduced FitzHugh-Nagumo model{, given by Eq.}~\eqref{eq:reldisp},
\begin{equation*}
\lambda^{(r)}_s = \mu+\frac{c}{\Lambda^s-\alpha c}+D_u \Lambda^s\quad\forall s=1,\dots N\, .
\end{equation*}
In particular, we are interested in determining the existence of negative $\Lambda^s$ for which $\lambda^{(r)}_s$ assumes a positive value. We thus consider the function $\lambda^{(r)}_s(x)$, for a generic variable $x$, and determine if there exists $x<0$ for which $\lambda^{(r)}_s(x)>0$. Consider first its derivative
\begin{equation*}
\frac{d\lambda^{(r)}_s}{dx} = -\frac{c}{(x-\alpha c)^2}+D_u \, ,
\end{equation*}
which vanishes for $x=\alpha c\pm\sqrt{\frac{c}{D_u}}$. One root, $x_+=\alpha c+\sqrt{\frac{c}{D_u}}$, is always positive, while the sign of the second, $x_-=\alpha c-\sqrt{\frac{c}{D_u}}$, depends on the parameters values. Finally, $\lambda^{(r)}_s(0)=\mu-1/\alpha<0$.

\begin{figure}[ht]
\centering
\includegraphics[scale=0.3]{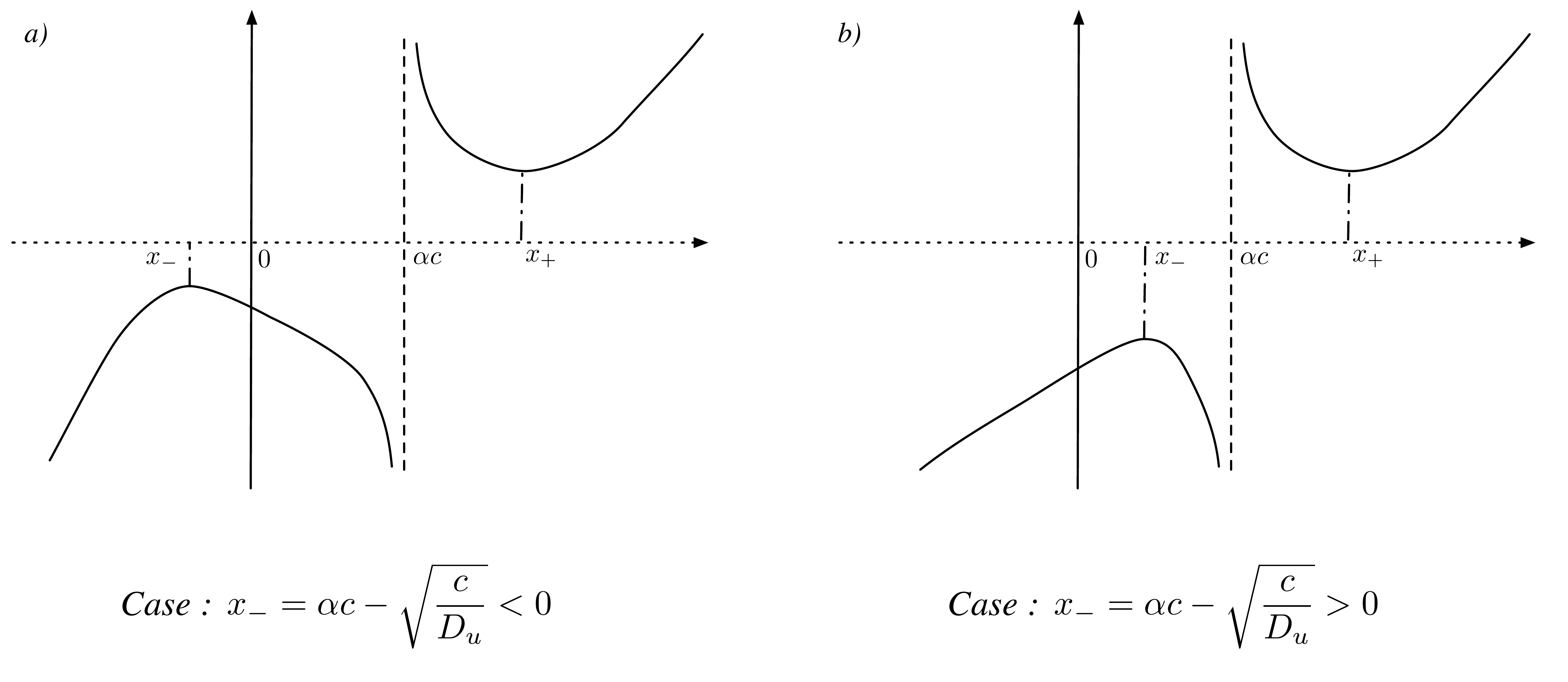}
\caption{\textbf{Qualitative behaviour of the dispersion relation for the {\em rFHN} model, $\lambda^{(r)}_s(x)$}. (a) The case $x_-=\alpha c-\sqrt{c/D_u}<0$, where Turing patterns may develop. (b) The case $x_-=\alpha c-\sqrt{c/D_u}>0$, where Turing patterns can never exist.}
\label{fig:lambdas}
\end{figure}

 In Fig.~\ref{fig:lambdas}, we report the qualitative behaviour of the graph of $\lambda^{(r)}_s(x)$ based on the information obtained from the first derivative and its values at $x=0$. {Note that} the graphs in both panels can be shifted vertically (with the constraint for the condition $\lambda^{(r)}(0)<0$), that is why the $x$-axis (dotted) is only indicatively shown. The function has a vertical asymptote at $x=\alpha c$, it diverge to $+\infty$ (resp. $-\infty$) as $x\rightarrow +\infty$ (resp. $x\rightarrow-\infty$) and it has a positive minimum located at $x=x_+>0$ and a maximum, located at $x=x_-$, whose position and value can be positive or negative.

Because the spectrum of the Laplace matrix, $\Lambda^s$, is non-positive, we are interested in studying further the case $x_-=\alpha c-\sqrt{c/D_u}<0$, being the only possibility to have Turing patterns. Let us thus assume $x_-=\alpha c-\sqrt{c/D_u}<0$ and compute $\lambda^{(r)}_s(x_-)$,
\begin{equation*}
\lambda^{(r)}_s(x_-) = \mu-2\sqrt{D_u c}+D_u \alpha c\, ,
\end{equation*}
the latter is positive if and only if
\begin{equation}
\mu>D_u c\left(\frac{2}{\sqrt{D_u c}}-\alpha\right)\, .
\end{equation}
The right hand side is positive because of the assumption $\alpha c-\sqrt{c/D_u}<0$.

\section{Dispersion {relations} of the original and reduced models}
\label{sec:reldispcomparison}

We hereby compare the dispersion relations of the {\em FHN} and {\em rFHN} models.
Substituting $\gamma=cD_v$ into Eq.~(\ref{eq:reldispFHNmain}) and considering $D_v\gg 1$, we can transform the dispersion relation for the {\em FHN} model as
\begin{eqnarray}
\label{eq:reldispFHNbis2}
\lambda_s&=&\frac{1}{2}\left[\mu -\alpha cD_v+\Lambda^s (D_u+D_v)+\sqrt{-4cD_v+\left[\mu+\alpha cD_v+(D_u-D_v)\Lambda^s\right]^2}\right]\notag\\
&=&\frac{1}{2}
\Big[
\mu -\alpha cD_v+\Lambda^s (D_u+D_v)+\left[\mu+\alpha cD_v+(D_u-D_v)\Lambda^s\right]
\cr
&&\times
\sqrt{1-\frac{4cD_v}{\left[\mu+\alpha cD_v+(D_u-D_v)\Lambda^s\right]^2}}
\Big]\notag\\
&=&\frac{1}{2}
\Big[
\mu -\alpha cD_v+\Lambda^s (D_u+D_v)+\left[\mu+\alpha cD_v+(D_u-D_v)\Lambda^s\right] 
\cr
&&\times
\left(1-\frac{1}{2}\frac{4cD_v}{\left[\mu+\alpha cD_v+(D_u-D_v)\Lambda^s\right]^2}+\mathcal{O}(1/D_v^2)\right) \ 
\Big]\notag\\
&=&\mu +\Lambda^s D_u+\frac{c}{\left(1-(D_u/D_v)\right)\Lambda^s-\alpha c -\mu/D_v}+\mathcal{O}(1/D_v)\notag\\
&=&\mu +\Lambda^s D_u+\frac{c}{\Lambda^s-\alpha c}+\mathcal{O}(1/D_v)=\lambda^{(r)}_s +\mathcal{O}(1/D_v)\, ,
\end{eqnarray}
where $\lambda_s^{(r)}$ is the dispersion relation for the reduced FHN model. Thus, by taking the limit $D_v\rightarrow \infty$, the claim in Eq.~(\ref{eq:limreldisp}) follows.

\section{The case of directed networks}
\label{sec:dirnet}

In the main text, we considered as underlying support an undirected network, a link connecting nodes $i$ and $j$ can be traveled in both direction by the species; however, in many relevant real cases, links among nodes result to be directed {or  the species cannot move with the same easiness in both directions}. The framework of directed network is thus more adequate to model such systems. In this case, the adjacency matrix satisfies $A_{ij}=1$ if there is a link from $i$ to $j$ and it is not in general a symmetric one. We define the out-degree, i.e. the number of links exiting from a node, by $k^{out}_i = \sum_j A_{ij}$ and the Laplace matrix $L_{ij} = A_{ij} - \delta_{ij} k^{out}_i$. Its spectrum is now in general complex except for the eigenvalue $\Lambda^1=0$ associated with the eigenvector $\boldsymbol{\phi}^1=(1,\dots,1)^T$.

\begin{figure}[ht]
\centering
\includegraphics[scale=0.25]{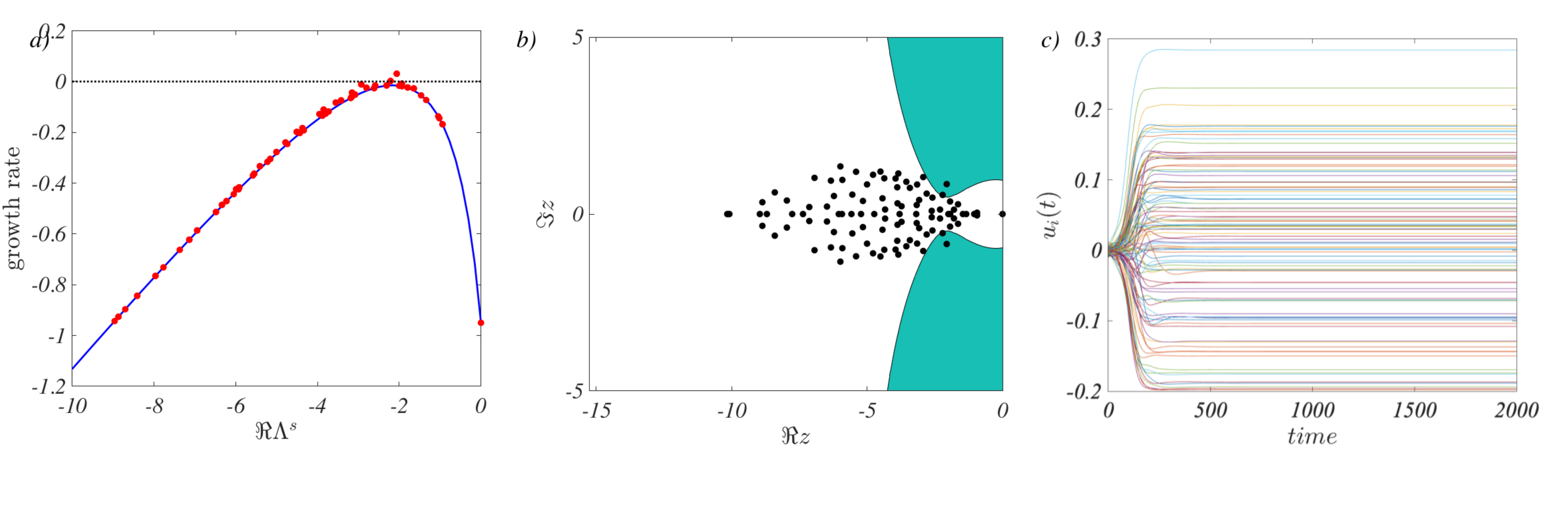}
\caption{\textbf{Turing patterns on a directed network.} {(a)} Dispersion relation for the continuous support (blue line) and for the networked system (red dots). We can observe that the blue line lies below zero, so no patterns can develop on the continuous domain or a symmetric network, {while} there are (few) red dots associated with a positive $\Re\lambda^s$ meaning that patterns can develop on a directed network. (b) Dispersion relation in the complex plane. The green region corresponds to $\Re\lambda^s>0$ and the complex spectrum of the Laplace matrix is represented by black dots. We can observe that some eigenvalues lie in the green region. The network is a directed  Erd\H{o}s-R\'enyi network made by $N=100$ nodes and probability for existence of a link $p=0.05$. {(c)} Time evolution of $u_i(t)$ for each node. The model parameters have been set to: $\mu=1.05$, $\beta=0$, $\alpha=0.5$, $c=2$ and $D_u=0.2$.}
\label{fig:utpattdir}
\end{figure}

The equation ruling the time evolution of species concentration is formally identical to Eq.~\eqref{eq:fhnmodel2} except for the use of the directed form of the Laplace matrix. One can thus perform the same stability analysis of the homogeneous solution $\hat{u}=0$ under heterogeneous perturbation and end up with the following dispersion relation:
\begin{equation}
\label{eq:reldispcmpx}
\Re\lambda_s = \mu+\frac{c(\Re\Lambda^s-\alpha c)}{(\Re\Lambda^s-\alpha c)^2+(\Im\Lambda^s)^2}+D_u \Re\Lambda^s\quad\forall s=1,\dots n\, ,
\end{equation}
where we explicitly wrote the dependence on the real and imaginary part of $\Lambda^s$. In Fig.~\ref{fig:utpattdir}, we report the dispersion relation (a) for a set of parameters for which Turing patterns do emerge on a directed support but not on a symmetric one. Indeed, the blue curve (associated with a continuous support or a symmetric network) always lies below the zero level. The panel (b) shows a different view of the dispersion relation; in the complex plane $(\Re\Lambda,\Im\Lambda)$, we draw the region (green) where $\Re\lambda^s$ is positive and the spectrum $\Lambda^s$ (black dots). Whenever an eigenvalue lies in the former region, it determines a positive dispersion relation and thus the emergence of patterns. The {typical patterns obtained by numerical simulations} are shown on the panel (c).

Let us observe that the study hereby presented is similar to the one performed in Ref.~\cite{asllaniNatComm}. However, the dimensionality {reduction} we can achieve using the adiabatic elimination makes simpler the definition of the domain of instability.

\section{A general model}
\label{sec:genmodel}
We have shown that the {\em FHN} model and the reduced one do exhibit Turing patterns for the same set of parameters. The mathematical reason {for this is} that they {share} the same roots for the dispersion relation and indeed this is due to the fact that the {dispersion relation} does not depend on the parameter $D_v$. In fact, the roots {of} Eqs.~\eqref{eq:rootreldispFHNmain} are obtained {by} setting the {following term in Eq.~\eqref{eq:reldispFHNmain} to zero,}
\begin{equation*}
\left[\gamma(1-\alpha \mu)+\Lambda^s(\mu D_v-\gamma\alpha D_u)+D_uD_v(\Lambda^s)^2\right]=0\, ,
\end{equation*}
{which gives, by} replacing $\gamma=cD_v$,
\begin{equation*}
\left[cD_v(1-\alpha \mu)+\Lambda^s(\mu D_v-cD_v\alpha D_u)+D_uD_v(\Lambda^s)^2\right]=0\, .
\end{equation*}
We can hence factor out $D_v$ {as}
\begin{equation*}
D_v\left[c(1-\alpha \mu)+\Lambda^s(\mu-c\alpha D_u)+D_u(\Lambda^s)^2\right]=0\, ,
\end{equation*}
from which we can conclude that the roots are independent from $D_v$.

A natural question is thus to understand how general is this result. We are thus consider a generic two-species reaction-diffusion model defined on a network,
\begin{equation}
\begin{cases}
\dot u_i &=F(u_i,v_i)+D_u \sum_j L_{ij}u_j\\
\dot v_i &=G(u_i,v_i)+D_v \sum_j L_{ij}v_j\, ,
\end{cases}
\label{eq:gensyst}
\end{equation}
and {perform} the adiabatic elimination {by} letting $D_v\rightarrow \infty$. We thus assume, after properly rescaling the parameters involved in $G$, that
\begin{equation}
\lim_{D_v\rightarrow \infty}\frac{G(u_i,v_i)}{D_v}=g(u_i,v_i)\, ,
\label{eq:cond1}
\end{equation}
for some smooth function $g$. We moreover assume the zeros of $g$ to lie on a graph {of} $v$ versus $u$, that is
\begin{equation}
g(u_i,v_i)=g_1(u_i)-av_i\, ,
\label{eq:cond2}
\end{equation}
for some $g_1$ and $a> 0$.

Under the previous assumptions, in the limit $D_v\rightarrow \infty$ the second equation of~\eqref{eq:gensyst} returns
\begin{equation}
0 = g_1(u_i)-av_i+ \sum_j L_{ij}v_j\quad \forall i=1,\dots,N\, ,
\label{eq:limit}
\end{equation}
and, defining $\mathbf{M}=(\mathbf{L}-a\mathbf{I})^{-1}$, we obtain
\begin{equation}
v_i = \sum_j M_{ij}g_1(u_j)\quad \forall i=1,\dots,N\, .
\label{eq:forv}
\end{equation}
We can finally insert this relation into the remaining equation of~\eqref{eq:gensyst} to derive the time evolution of the reduced model:
\begin{equation}
\partial_t u_i =F\left(u_i,\sum_j M_{ij}g_1(u_j)\right)+D_u \sum_j L_{ij}u_j\, .
\label{eq:redmod}
\end{equation}

Let us denote by $(u_i,v_i)=(\bar{u},\bar{v})$ the homogeneous equilibrium of the original system, namely $F(\bar{u},\bar{v})=G(\bar{u},\bar{v})=0$. To ensure that the reduced model has the same equilibrium, we have to impose
\begin{equation*}
\bar{v} = \sum_j M_{ij}g_1(\bar{u})\, ,
\end{equation*}
that is 
\begin{equation}
\bar{v} = \frac{g_1(\bar{u})}{a}\, .
\label{eq:constraint}
\end{equation}
Stated differently, $G$ and $g$ {should} have the same zeros for all values of $D_v$.

The roots of the dispersion relation can be obtained by imposing 
\begin{equation*}
\left[\partial_u F\partial_v G-\partial_uG \partial_vF+\Lambda^s(D_u \partial_v G+D_v \partial_u F)+D_uD_v(\Lambda^s)^2\right]=0\, ,
\end{equation*}
{and, in order} to be able to factor out $D_v$ again and based on the above assumption on $G$, we must require $G(u,v)=D_v \left[g(u)-av\right]$. The {\em FHN} {model} is thus a particular case of this class of systems.

\begin{remark}[Reduced models in generic slow-fast systems]
Let us observe that the procedure presented in this work can be generalised to other reaction models and in particular cast into the two-species slow-fast reaction-diffusion system defined on a network
\begin{equation*}
\begin{cases}
\dot u_i &=f(u_i)+g(v_i)+D_u \sum_j L_{ij}u_j\\
\epsilon \dot v_i &=h(u_i)-v_i+D_v \sum_j L_{ij}v_j\, ,
\end{cases}
\end{equation*}
where $u_i$ acts as an activator while $v_i$ is an inhibitor.

Taking the limit $\epsilon\rightarrow 0$ and adiabatically eliminating $v_i$ gives
\begin{equation*}
0=h(u_i)-v_i+D_v \sum_j L_{ij}v_j\, ,
\end{equation*}
hence $\sum_j (\delta_{ij}-D_v L_{ij})v_j=h(u_i)$, or in matrix form
\begin{equation*}
\mathbf{M}\vec{v}:= (\mathbf{I}-D_v \mathbf{L})\vec{v}=\vec{h}(u)\, ,
\end{equation*}
with $\vec{h}(u)=({h}(u_1),\dots,{h}(u_N))$ and where $\mathbf{M}$ is defined by the left hand side.

Eventually, we obtain the reduced model by inserting the previous relation in the remaining equation describing the evolution of the activator in each node, $u_i${, as}
\begin{equation*}
\dot u_i =f(u_i)+g\left(\sum_jM_{ij} h(u_j)\right)+D_u \sum_j L_{ij}u_j\, .
\end{equation*}
Let us observe that in the case of reaction-diffusion {equations} on a continuum medium, we obtain a nonlocal kernel as a solution to the Helmholtz equation in place of $\mathbf{M}$. 
\end{remark}

\end{document}